\documentclass[pss]{wiley2sp} % provides new 2008 pss two-column style (no alternative manuscript style output available at present)
\usepackage{amsmath}
\begin{document}

\title{Magnetic Gr\"{u}neisen ratio of the random transverse-field Ising chain}
\titlerunning{Magnetic Gr\"{u}neisen ratio of the random transverse-field Ising chain}

\author{%
  Thomas Vojta\textsuperscript{\Ast,\textsf{\bfseries 1},\textsf{\bfseries 2}},
  Jos\'{e} A. Hoyos\textsuperscript{\textsf{\bfseries 3}}
}
% Abbreviated list of authors for the page headers
\authorrunning{T. Vojta and J.~A. Hoyos}

%E-mail-address of corresponding author
\mail{e-mail  \textsf{vojtat@mst.edu}, Phone:  +1-573-341 4793, Fax: +1-573-341 4715}

% author's affiliations/addresses
\institute{%
  \textsuperscript{1}\,Department of Physics, Missouri University of Science and Technology, Rolla, MO 65409,
  USA\\
  \textsuperscript{2}\,Max-Planck-Institute for Physics of Complex Systems, Noethnitzer Str. 38, 01187 Dresden,
  Germany\\
  \textsuperscript{3}\,Department of Physics, Duke University, Durham, NC 27708, USA}

\received{XXXX, revised XXXX, accepted XXXX} % do not change, will be filled in by the publisher
\published{XXXX} % do not change, will be filled in by the publisher

%Please select four to six PACS-codes from the enclosed list (PACS.txt) or from www.aip.org/pacs)
\pacs{71.27.+a, 75.10.Nr, 75.40.Cx, 71.10.Hf} % For example: 71.20.Ps

\abstract{%
%
% Usage: \abstcol{<left column>}{<right column>}
\abstcol{%
  The magnetic analog of the Gr\"{u}neisen parameter, i.e., the magnetocaloric effect,
  is a valuable tool for studying field-tuned quantum phase transitions. We determine
  the magnetic Gr\"{u}neisen parameter of the one-dimensional random transverse-field
  Ising model, focusing on its low-temperature behavior at the exotic infinite-randomness
  quantum critical point and in the associated quantum Griffiths phases.
  }{%
  We present extensive numerical simulations showing that the magnetic Gr\"{u}neisen parameter
  diverges logarithmically with decreasing temperature in the quantum Griffiths phase. It changes
  sign right at criticality. These results confirm a recent strong-disorder renormalization group
  theory.
  We also compare our findings to the behavior of the clean transverse-field Ising chain.
  }}

\maketitle   % please do not remove

%%%%%%%%%%%%%%%%%%%%%%%%%%%%%%%%%%%%%%%%%%%%%%%%%%%%%%%%%%%%%%%%%%%%
% Main text starts here!
%%%%%%%%%%%%%%%%%%%%%%%%%%%%%%%%%%%%%%%%%%%%%%%%%%%%%%%%%%%%%%%%%%%%
\section{Introduction}
\label{Sec:Intro}
%%%%%%%%%%%%%%%%%%%%%%%%%%%%%%%%%%%%%%%%%%%%%%%%%%%%%%%%%%%%%%%%%%%%

Zero-temperature quantum phase transitions are characterized by a competition
between different quantum ground states. The resulting quantum fluctuations
often cause
unconventional behavior of thermodynamic and transport properties in the
vicinity of the phase transition. Celebrated consequences include non-Fermi liquid
behavior in metals and the emergence of new phases such as exotic superconductors
(for recent reviews see, e.g., Refs.\
\cite{SGCS97,Sachdev_book99,Vojta_review00,VojtaM03}).

Realistic systems often feature considerable amounts of quenched randomness in the form
of vacancies, impurities, or other defects. In these systems, the combination of
static random fluctuations and quantum fluctuations leads to dramatic disorder effects
in the vicinity of a quantum phase transition. They include quantum Griffiths
singularities \cite{ThillHuse95,RiegerYoung96},
infinite-randomness critical points with activated dynamical scaling
\cite{Fisher92,Fisher95},
and smeared quantum phase transitions \cite{Vojta03a,HoyosVojta08}.
A recent review of these phenomena can be found in Ref.\ \cite{Vojta06}.

In recent years, the Gr\"{u}neisen parameter $\Gamma$ \cite{Grueneisen12,ZGRS03}
has been established as a
valuable tool for characterizing pressure-tuned quantum phase transitions. It is defined
as the ratio between the thermal volume expansion coefficient $\beta$ and the specific heat
$c_p$. If the quantum phase transition is tuned by varying an external magnetic field,
one needs to consider the magnetic analog $\Gamma_H$ of the Gr\"{u}neisen parameter, defined as
the ratio between the temperature-derivative of the magnetization and the specific heat
$c_H$. It can be directly measured via the magnetocaloric effect.

According to Zhu et al.\ \cite{ZGRS03,GarstRosch05}, the thermal expansion coefficient
is more singular than the specific heat at a generic pressure-tuned quantum critical point,
leading to a diverenge of the Gr\"{u}neisen parameter. In particular, these authors showed
that $\Gamma \sim T^{-1/(z\nu)}$ if the temperature $T$ is lowered at the critical pressure
$p_c$ and $\Gamma \sim 1/(p-p_c)$ if the pressure $p$ approaches its critical value at
zero temperature. Here, $z$ and $\nu$ are the dynamical and correlation length critical
exponents,  respectively. These results hold below the upper critical dimension
(where hyperscaling is valid). Above the upper critical dimension,
$\Gamma$ still diverges, but the functional form depends on dangerously irrelevant
variables. The divergence of the Gr\"{u}neisen parameter has been confirmed experimentally
at several pressure-tuned quantum phase transitions, and the functional form
of the divergence can be used to discriminate between different theories
\cite{Kuechleretal03,Kuechleretal04,Kuechleretal06,TRGSG09}.

The above results apply to clean quantum critical points with conventional power-law scaling.
Given the utility of the Gr\"{u}neisen parameter, it is very desirable to know its behavior
within the unconventional quantum phase transition scenarios emerging in the presence of quenched randomness.
Recently, the low-temperature behavior of Gr\"{u}eneisen parameter $\Gamma$
and its magnetic analog $\Gamma_H$ were studied at infinite-randomness quantum critical
points and in the associated quantum Griffiths phases \cite{Vojta09}. Using heuristic rare
region arguments as well as a scaling theory, it was predicted that the Gr\"{u}neisen parameter
diverges logarithmically with decreasing temperature $T$ in the quantum Griffiths phases
and as a higher power of $\ln(1/T)$ right at the infinite-randomness quantum critical point.

In this paper, we confirm and illustrate these predictions by considering a specific
microscopic model featuring an infinite-randomness quantum critical point, viz.,
the random transverse-field Ising chain. Since the transition is tuned by magnetic field,
we study the magnetic analog $\Gamma_H$ of the Gr\"{u}neisen parameter (i.e., the
magnetocaloric effect). Our paper is organized as follows.
In Sec.\ \ref{Sec:Model}, we define the model, and in Sec.\ \ref{Sec:Gamma_H}, we
briefly summarize the results of the scaling theory of Ref.\ \cite{Vojta09}. Section
\ref{Sec:Simulations} is devoted to extensive
numerical calculations of the magnetic Gr\"{u}neisen parameter. We conclude in Sec.\
\ref{Sec:Conclusions}.

%%%%%%%%%%%%%%%%%%%%%%%%%%%%%%%%%%%%%%%%%%%%%%%%%%%%%%%%%%%%%%%%%%%%
\section{Random transverse-field Ising chain}
\label{Sec:Model}
%%%%%%%%%%%%%%%%%%%%%%%%%%%%%%%%%%%%%%%%%%%%%%%%%%%%%%%%%%%%%%%%%%%%

We consider a one-dimensional random transverse-field Ising model given
by the Hamiltonian
\begin{equation}
H=  -  \sum_{i=1}^{L_0}J_{i}\sigma_{i}^{z}\sigma_{i+1}^{z}-\sum_{i=1}^{L_0}h_{i}\sigma_{i}^{x}~.
\label{eq:H}
\end{equation}
Here, $\sigma_{i}^{x}$ and $\sigma_{i}^{z}$ are the Pauli matrices representing the
$x$ and $z$ components of the spin-1/2 at site $i$, and $L_0$ is the chain length.
The bonds $J_{i}>0$ and transverse fields $h_{i}>0$ are independent random variables with
bare (initial) probability distributions $P_I(J)$ and $R_I(h)$.
The physics of this model has been studied extensively in the literature.
If the bonds are sufficiently strong compared to the fields, the ground state is
ferromagnetic, with a spontaneous magnetization in $\pm z$ direction. In the opposite limit,
for sufficiently strong fields, the ground state is paramagnetic with vanishing
$z$-magnetization. The two ground-state phases are separated by a quantum critical point
located at $[\ln(J_i)]_{\rm dis} = [\ln(h_i)]_{\rm dis}$ where $[\ldots]_{\rm dis}$
denotes the disorder average. The model does not show long-range order at any
nonzero temperature.

The quantum critical behavior can be determined exactly
by means of the strong-disorder renormalization group \cite{Fisher92,Fisher95}.
It does not depend on the details of the distributions $P_I(J)$ and $R_I(h)$
provided they are not too singular. The quantum critical fixed point
that emerges from this analysis is of exotic infinite-randomness type.
This implies that correlation time $\xi_\tau$ and correlation
length $\xi$ are related by the exponential law $\ln(\xi_\tau) \sim \xi^\psi$ with a
tunneling exponent $\psi=1/2$. In contrast, the relation between the correlation length and the
distance $r$ from criticality
%$r \sim [\ln(h_i)]_{\rm dis} -[\ln(J_i)]_{\rm dis}$
is conventional, $\xi \sim |r|^{-\nu}$, with a correlation length exponent $\nu=2$.

In the following, we consider systems with randomness in the bonds $J_i$ only.
The transverse magnetic field is taken to be uniform,
$R_I(h) = \delta (h-H)$ because we wish to view it as the experimental
tuning parameter.

%%%%%%%%%%%%%%%%%%%%%%%%%%%%%%%%%%%%%%%%%%%%%%%%%%%%%%%%%%%%%%%%%%%%
\section{Magnetic Gr\"{u}neisen parameter}
\label{Sec:Gamma_H}
%%%%%%%%%%%%%%%%%%%%%%%%%%%%%%%%%%%%%%%%%%%%%%%%%%%%%%%%%%%%%%%%%%%%

The magnetic analog $\Gamma_H$ of the Gr\"{u}neisen parameter is defined as the ratio
of the temperature derivative of the magnetization
$M=\sum_i \langle \sigma_i^x \rangle$
and the specific heat, both taken at constant field $H$ \cite{ZGRS03}.
Note that $M$ is the $x$-magnetization (the magnetization conjugate to the external
field)
rather than the $z$-magnetization (the order parameter of the transition).
Because the order parameter vanishes at any nonzero temperature, $M$
is the magnetization measured in experiment. We thus define
\begin{equation}
\Gamma_H = -\frac {(\partial M/\partial T)_H} {C_H} = -\frac {(\partial S/\partial H)_T}{T(\partial S/\partial T)_H}
         = \frac 1 T \left(\frac {\partial T}{\partial H}\right)_S
\label{eq:Gamma_H_def}
\end{equation}
with $S$ the entropy. Here, $(\partial S / \partial H)_T$ explores the dependence of entropy
on the dimensionless distance from criticality, $r=(H-H_c)/H_c$.
\footnote{Within the strong-disorder renormalization group \cite{Fisher92,Fisher95},
the distance from criticality is naturally measured in terms of
$\delta \sim [\ln(h_i)]_{\rm dis} -[\ln(J_i)]_{\rm dis}$. Close to criticality,
$r$ and $\delta$ are proportional to each other. Using $\delta$ instead of $r$
would thus introduce some constant factors at intermediate stages of the calculations,
but leave the final results unchanged.}
The last equality in (\ref{eq:Gamma_H_def})
shows that $\Gamma_H$ can be obtained directly from the magnetocaloric effect.

Let us now apply the scaling theory \cite{Vojta09} for the magnetic
Gr\"{u}neisen parameter to our case. According to the strong-disorder renormalization group
\cite{Fisher92,Fisher95}, the entropy is proportional to the number of independent
active clusters at energy scale $T$. It thus has the scaling form
(see, e.g., Ref.\ \cite{VojtaKotabageHoyos09})
\begin{equation}
S(r,T) = L_0 s_0 [\ln(T_0/T)]^{-2} \, \Phi\left (r\,\ln(T_0/T)\right)
\label{eq:S_scaling}
\end{equation}
with $s_0 =\ln(2)$ the entropy (in units of $k_B$) of a single cluster and $T_0$ a microscopic
temperature scale. The scaling function
$\Phi(y)$ is analytic because there is no phase transition at any finite
temperatures. For small $y$, we therefore expand
$\Phi(y) =\Phi(0) + y\Phi^\prime (0) + \frac 1 2 y^2 \Phi^{\prime \prime}(0) + \ldots $.
At the quantum critical point, the random transverse-field Ising model is self-dual
(invariant under the exchange $h_i \leftrightarrow J_i$). The scaling function is
therefore even in $y$ which implies $\Phi^\prime(0)=0$.
For large $y$, i.e., in the Griffiths phase, $\Phi(y) = A y^2 \exp(-c |y|)$ with
$A$ and $c$ constants.

We first consider the quantum critical region defined by $|r| \ln(T_0/T) \ll 1$. Taking
the appropriate derivatives of the entropy and keeping only the leading terms, we find
the specific heat
\begin{equation}
C_H = 2 L_0 s_0 \Phi(0) [\ln(T_0/T)]^{-3}
\end{equation}
and the magnetization derivative
\begin{equation}
\left ( {\partial M }/ {\partial T} \right)_H = L_0 s_0 \, r \Phi^{\prime \prime}(0) /
H_c.
\end{equation}
This results in a magnetic Gr\"{u}neisen parameter
\begin{equation}
\Gamma_H = - \frac {H-H_c} {2 H_c^2} [\ln(T_0/T)]^3 \frac {\Phi^{\prime\prime}(0)}{\Phi(0)}~
\label{eq:Grueneisen-critical}
\end{equation}
which has a sign change right at criticality $r=0$. Note that $\Gamma_H$ takes a
different form than in the generic theory of Ref.\ \cite{Vojta09} because
the first derivative of the scaling function vanishes in our system.

In the quantum Griffiths phase, $|r| \ln(T_0/T) \gg 1 $, we use the large-argument limit of
the scaling function $\Phi(y)$ to obtain the entropy $S=L_0 g(r) (T/T_0)^{\lambda(r)}$ with
$g(r)= A s_0 r^2$. The Griffiths exponent is given by $\lambda(r) =c |r|$. Taking
the appropriate derivatives leads to the specific heat
\begin{equation}
C_H = L_0 g(r)\, c|r| (T/T_0)^{\lambda(r)}
\end{equation}
and the magnetization derivative
\begin{equation}
(\partial M /\partial T)_H = - \frac {L_0}{H_c} g(r)\,  \frac{cr}{|r|} (T/T_0)^{\lambda(r)} \ln(T_0/T)~.
\end{equation}
Forming the ratio results in the magnetic Gr\"{u}neisen parameter
\begin{equation}
\Gamma_H = \frac 1 {H-H_c} \ln(T_0/T)~.
\label{eq:Grueneisen-griffiths}
\end{equation}
The Gr\"{u}neisen parameter thus diverges logarithmically with decreasing
temperature in the entire quantum Griffiths phase. The prefactor diverges
itself as $1/(H-H_c)$ when approaching criticality.

%%%%%%%%%%%%%%%%%%%%%%%%%%%%%%%%%%%%%%%%%%%%%%%%%%%%%%%%%%%%%%%%%%%%
\section{Computer simulations}
\label{Sec:Simulations}
%%%%%%%%%%%%%%%%%%%%%%%%%%%%%%%%%%%%%%%%%%%%%%%%%%%%%%%%%%%%%%%%%%%%

In this section, we report the results of extensive computer simulations of the
random transverse-field Ising chain (\ref{eq:H}) by means of a numerical implementation
of the strong-disorder renormalization group. We have studied long chains of up to
$10^6$ sites with periodic boundary conditions. All results are averages
over a large number ($N= 5000 \ldots 25000$) of disorder realizations.
The (bare) system is moderately
disordered; the bonds $J_i$ are drawn from the interval $[0.5,1]$ with a flat
distribution. The transverse field is uniform, $h_i \equiv H$. The critical field
can be determined analytically from the condition $[\ln(J_i)]_{\rm dis} = [\ln(h_i)]_{\rm
dis}$. This leads to $H_c=\exp[\ln(2)-1] \approx 0.73576$.

\begin{figure}[t]
\includegraphics[width=\linewidth,clip]{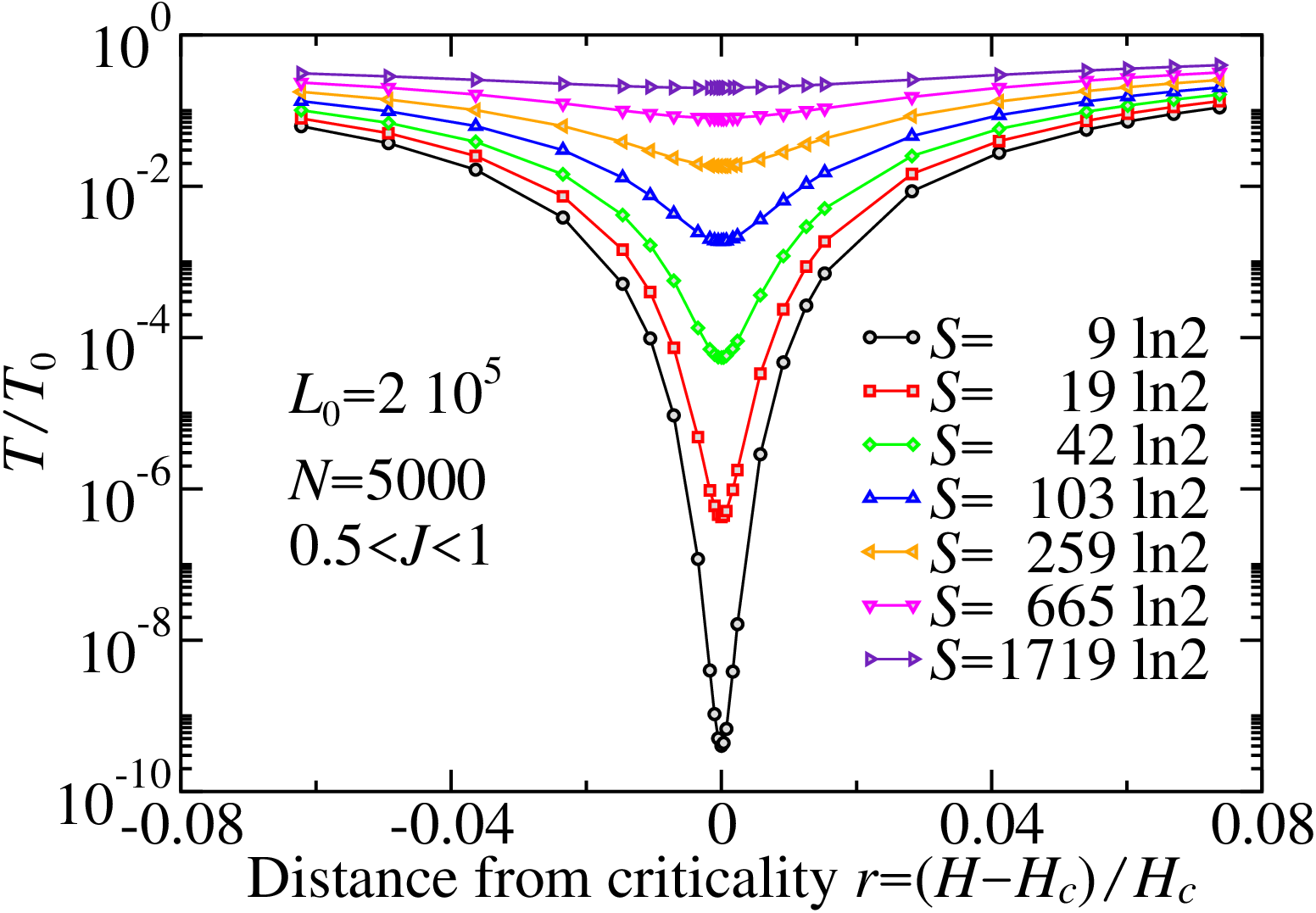}
\caption{Isentropes of the random transverse-field Ising model (\ref{eq:H}). Note the
symmetry of the cures w.r.t. $r \leftrightarrow -r$.}
\label{Fig:isentropes}
\end{figure}

\begin{figure}[t]
\includegraphics[width=\linewidth,clip]{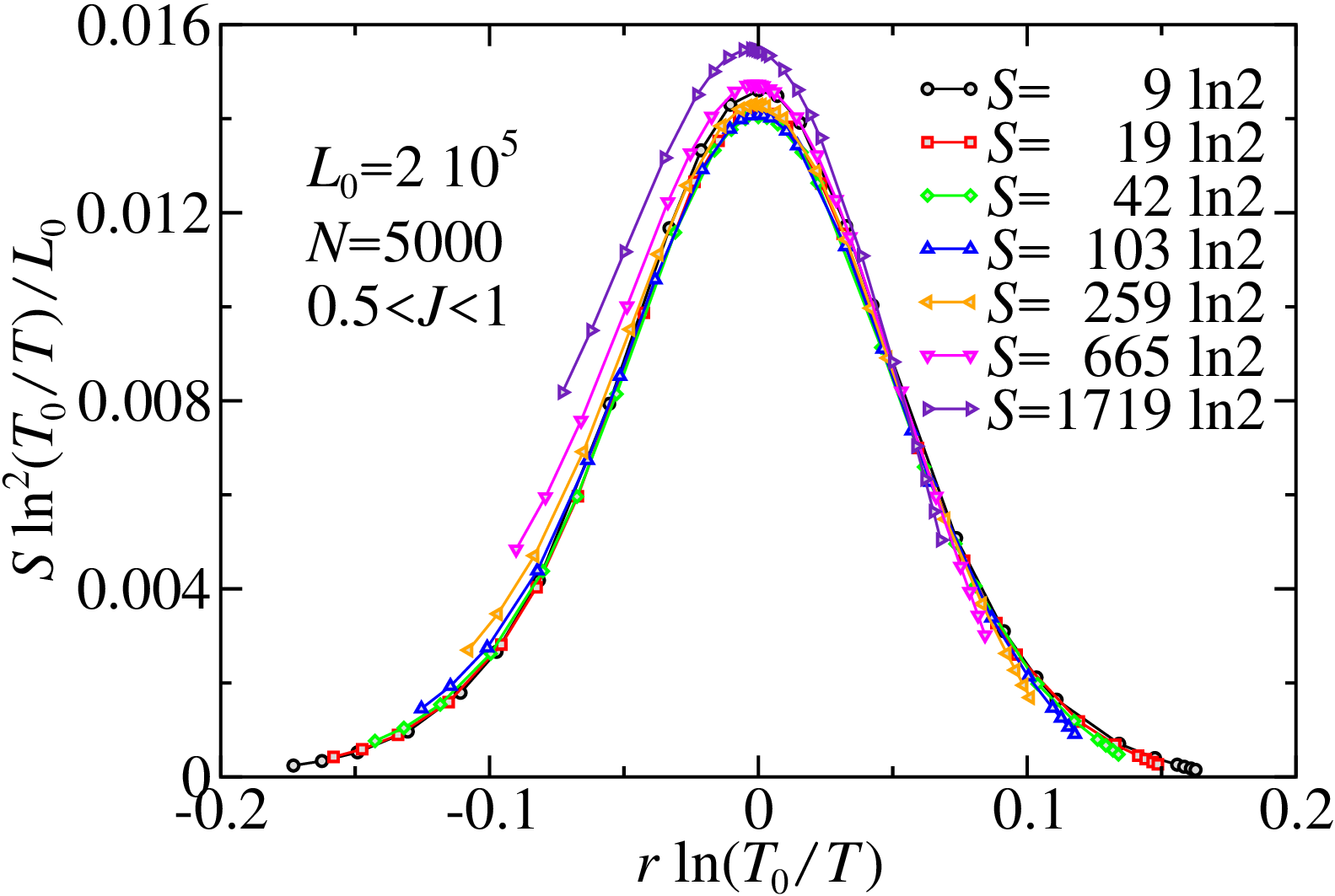}
\caption{Scaling plot of the isentropes of Fig.\ \ref{Fig:isentropes} according to the scaling form
(\ref{eq:S_scaling}) of the entropy.}
\label{Fig:isentropes-scaling}
\end{figure}

We first consider the behavior of the entropy $S$ as a function of temperature $T$
and distance to criticality $r$. To find the isentropes (curves of constant entropy),
we run the renormalization group until a certain fixed number $S/\ln(2)$ of clusters
is left. The corresponding temperature is then given by the energy cutoff of the
renormalization group.  Figure \ref{Fig:isentropes} shows the
resulting isentropes in the temperature-field plane.
The entropy strongly accumulates near the critical field. The derivative $(\partial S /\partial H)_T$
is thus positive on the magnetic side of the transition ($H<H_c$) and negative on the
nonmagnetic side ($H>H_c$). This is responsible for the
above-mentioned sign change of $\Gamma_H$ at criticality.
To verify the scaling form (\ref{eq:S_scaling}) of the entropy, we redraw the same data
in a scaling plot in Fig.\ \ref{Fig:isentropes-scaling}.
The figure shows that the high-entropy curves (which correspond to high temperatures and
the early stages of the renormalization group) do not scale perfectly and also reflect
the $H$-$J$ asymmetry of the bare system. However, with decreasing entropy (i.e.,
decreasing temperature), the scaling becomes better and better until it is almost perfect
within our error bars.

\begin{figure}
\includegraphics[width=\linewidth,clip]{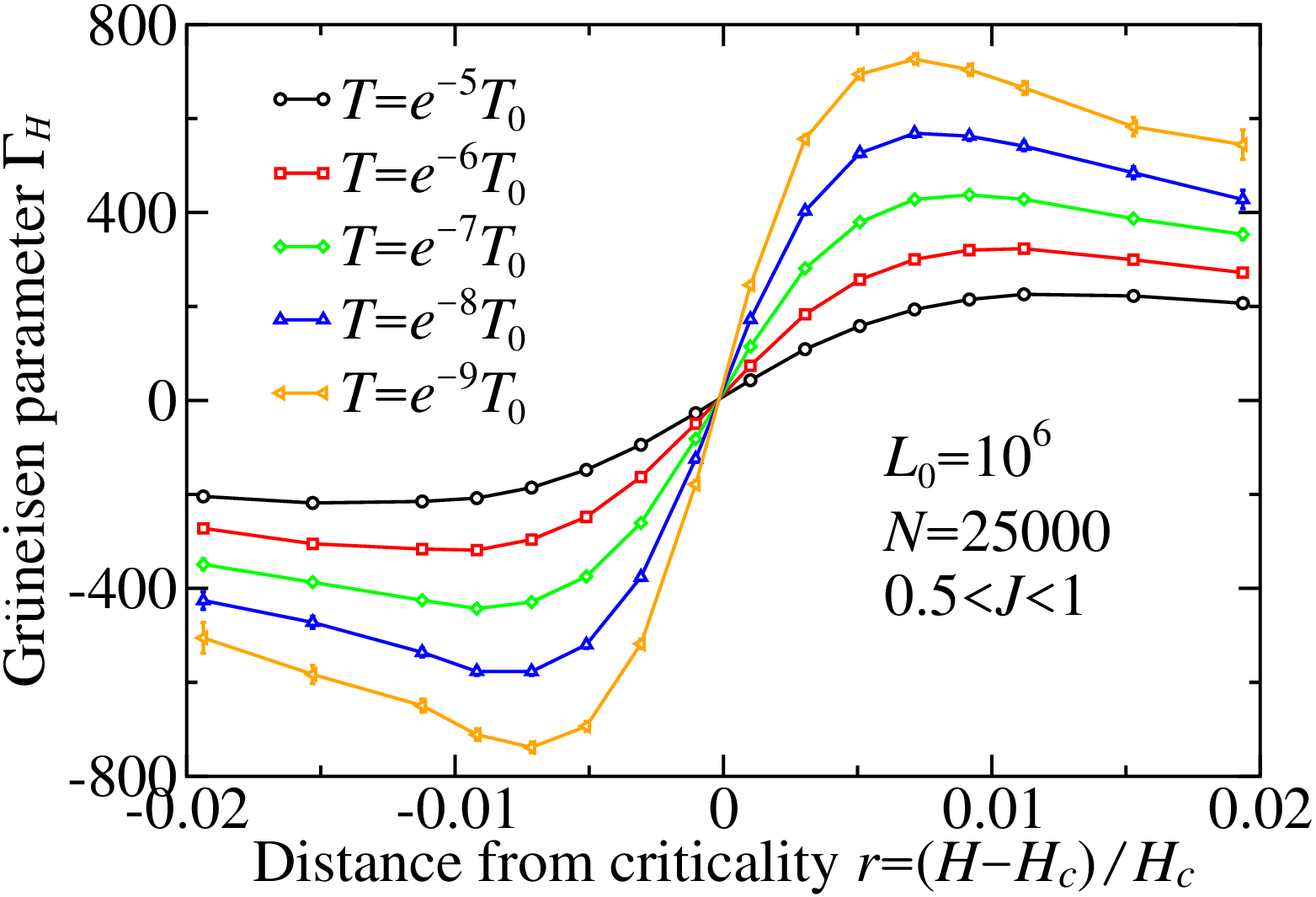}
\caption{Overview of the magnetic Gr\"{u}neisen parameter of the random transverse-field
Ising model.}
\label{Fig:Gruneisen-constanT}
\end{figure}

We now turn to the magnetic Gr\"{u}neisen parameter $\Gamma_H$. Figure
\ref{Fig:Gruneisen-constanT} gives an overview of $\Gamma_H$ as a function
of transverse field for various temperatures.
In agreement with the theory of Sec.\ \ref{Sec:Gamma_H}, the Gr\"{u}neisen parameter is
negative for $H<H_c$ and positive for $H>H_c$. At the critical field $H_c$,
it vanishes for all temperatures. The figures also shows that, at a given field $H$,
the magnitude of the Gr\"{u}neisen parameter increases with decreasing temperature.

\begin{figure}
\includegraphics[width=\linewidth,clip]{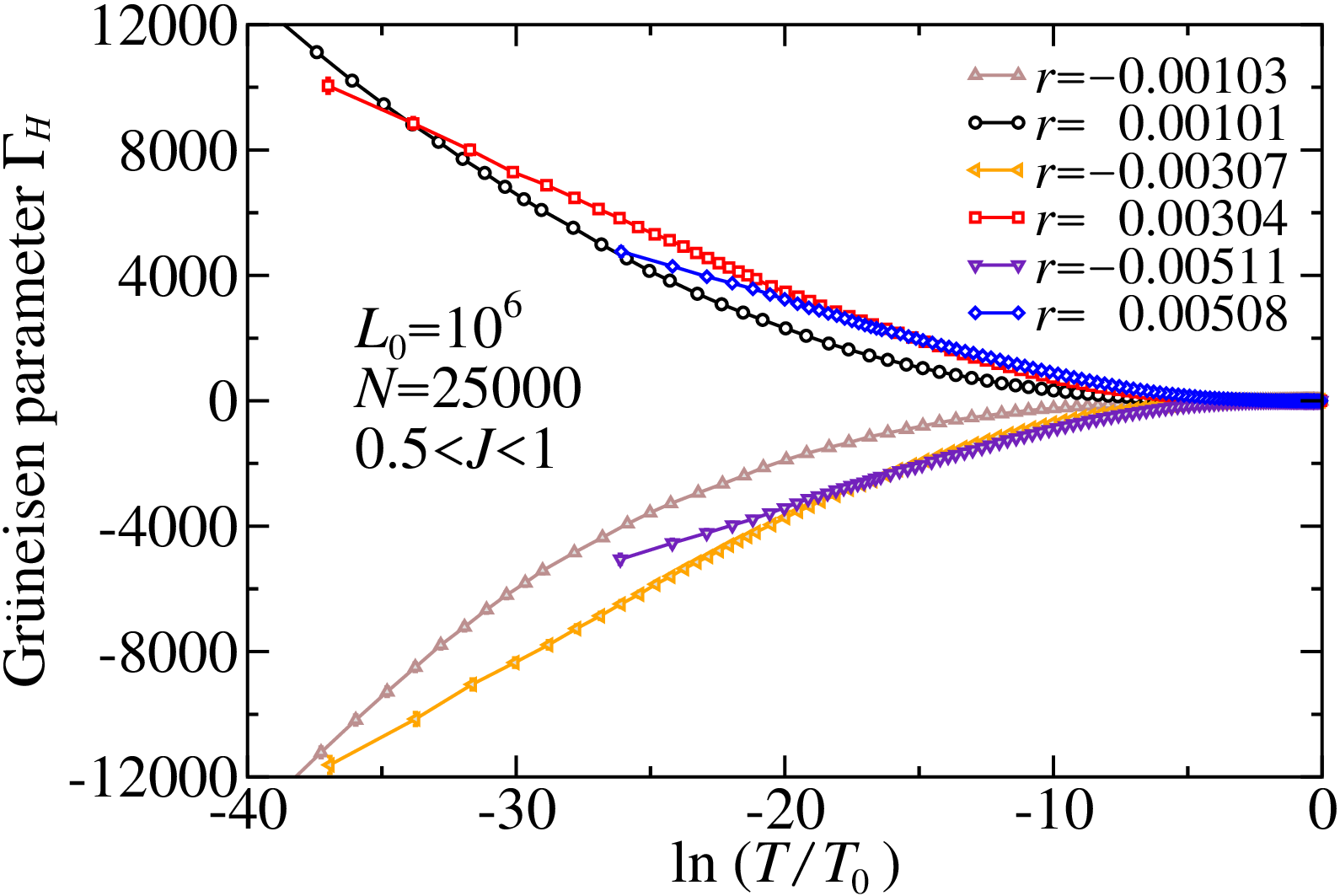}
\caption{Magnetic Gr\"{u}neisen parameter as a function of $\ln(T_0/T)$ for various
 values of the transverse field.}
\label{Fig:Gruneisen}
\end{figure}

To investigate more closely the temperature dependence of the Gr\"{u}neisen parameter
in the quantum Griffiths phase,
we plot $\Gamma_H$ as a function of $\ln(T/T_0)$ in Fig.\ \ref{Fig:Gruneisen}.
In the systems furthest away from criticality ($r=0.00508,-0.00511$), $\Gamma_H$
follows (after a very short high-temperature transient)
the $\ln(T_0/T)$ behavior predicted in (\ref{eq:Grueneisen-griffiths})
for the Griffiths phase. If we move closer to criticality
($r=0.00304, -0.00307$), the high-temperature behavior of $\Gamma_H$ is clearly
faster than $\ln(T/T_0)$.
At lower temperatures, it crosses over to the Griffiths phase form $\ln(T_0/T)$.
Finally, the systems closest to criticality ($r=0.00101, -0.00103$) do not reach
the crossover to the Griffiths phase form within the temperature range covered
in the simulations.

\begin{figure}
\includegraphics[width=\linewidth,clip]{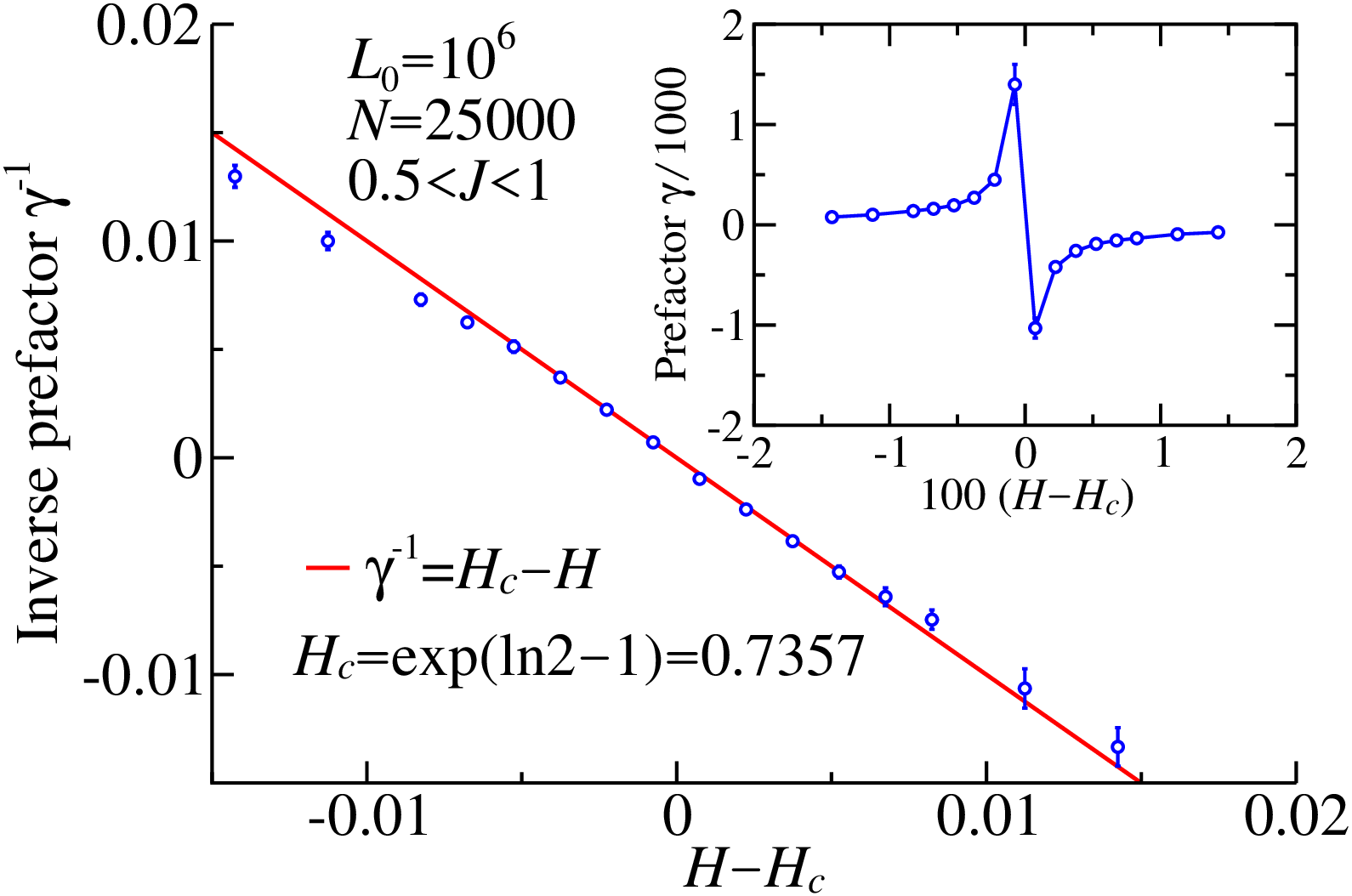}
\caption{Prefactors $\gamma =\Gamma_H / \ln(T_0/T)$ of the low-temperature Gr\"{u}neisen
parameter as a function of transverse field $H$.
The solid line represents the theoretical prediction from (\ref{eq:Grueneisen-griffiths})
without adjustable parameters.}
\label{Fig:prefactors}
\end{figure}

To confirm (\ref{eq:Grueneisen-griffiths}) quantitatively,
we have performed analogous calculations for many more values of the transverse field
and extracted the prefactors of the $\ln(T_0/T)$ low-temperature behavior of the
Gr\"{u}neisen parameter. These prefactors $\gamma = \Gamma_H / \ln(T_0/T)$ are plotted
in Fig.\ \ref{Fig:prefactors}.
The overview given in the inset shows that $\gamma$ increases and appears to diverge
close to criticality.
For a detailed comparison with (\ref{eq:Grueneisen-griffiths}), we plot
$\gamma^{-1}$ as a function of $H-H_c$ in the main panel of the figure. The numerical
data are in very good quantitative agreement with the prediction $\gamma^{-1} = H-H_c$.

\begin{figure}
\includegraphics[width=\linewidth,clip]{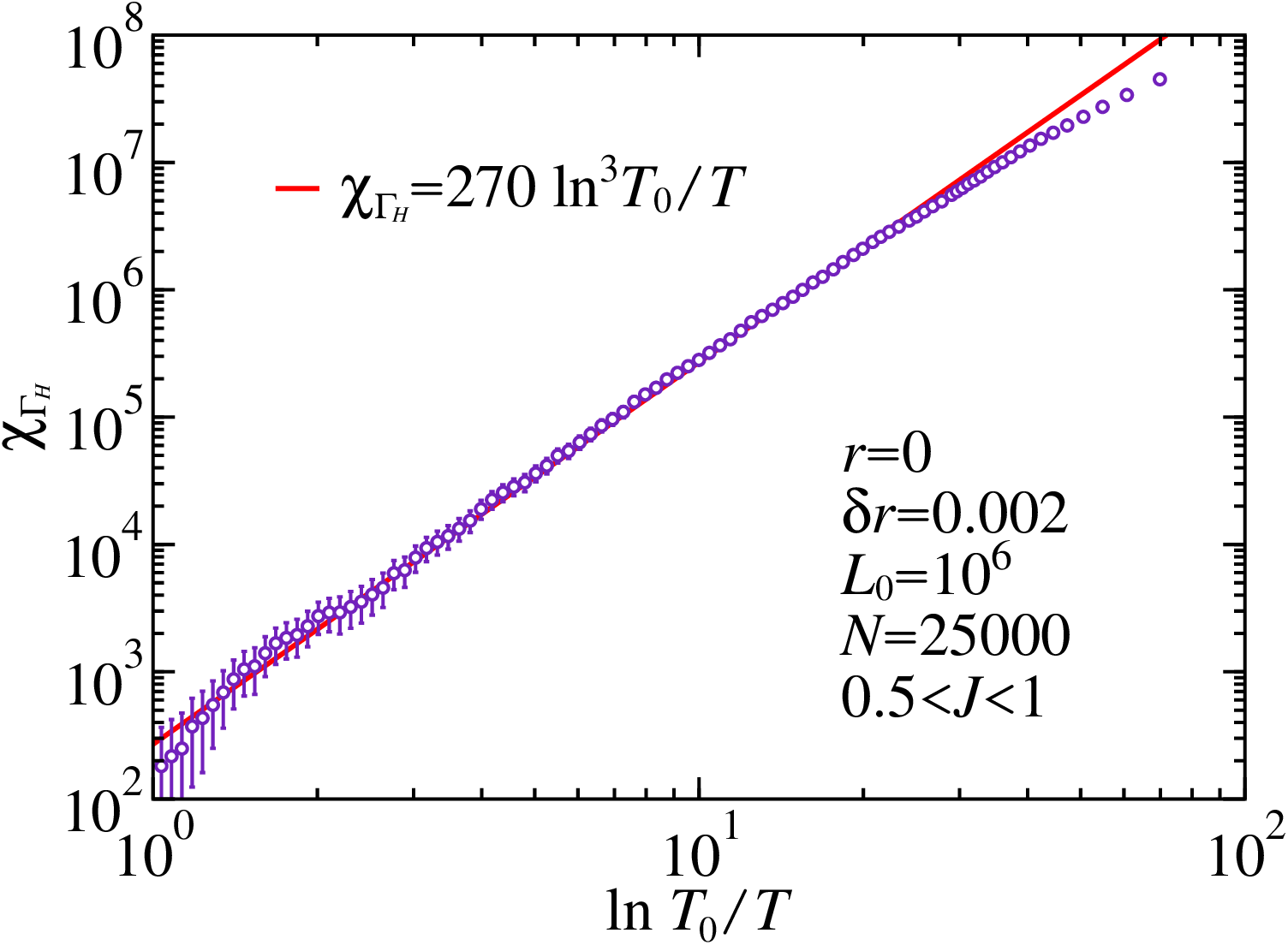}
\caption{Derivative $\chi_{\Gamma_H} = (\partial \Gamma_H /\partial r)_T$ of the Gr\"{u}neisen
        parameter taken at criticality, $r=0$. The solid line is a fit to the expected
        $[\ln(T_0/T)]^3$ behavior at low temperatures.}
\label{Fig:Grueneisen-susc}
\end{figure}

Finally, we turn to the behavior of the magnetic Gr\"{u}neisen parameter in the quantum
critical region, $|r| \ln(T_0/T) \ll  1$. To verify (\ref{eq:Grueneisen-critical}), we
compute the Gr\"{u}neisen ``susceptibility''
$\chi_{\Gamma_H} = (\partial \Gamma_H /\partial r)_T$ at criticality. According to
(\ref{eq:Grueneisen-critical}), this quantity is predicted to behave as $[\ln(T_0/T)]^3$
at low temperatures. The data in Fig.\ \ref{Fig:Grueneisen-susc}
show that $\chi_{\Gamma_H}$ indeed follows the expected behavior over many orders of
magnitude in $T$. The small deviations at the lowest temperatures can be attributed to
finite-size effects as well as systematic errors introduced by the finite field step
$\delta H$ in our simulations.

%%%%%%%%%%%%%%%%%%%%%%%%%%%%%%%%%%%%%%%%%%%%%%%%%%%%%%%%%%%%%%%%%%%%
\section{Conclusions}
\label{Sec:Conclusions}
%%%%%%%%%%%%%%%%%%%%%%%%%%%%%%%%%%%%%%%%%%%%%%%%%%%%%%%%%%%%%%%%%%%%

In summary, we have performed extensive computer simulations of the one-dimensional
random transverse-field Ising model by means of a numerical strong-disorder
renormalization group. We have studied the magnetic analog of the Gr\"{u}neisen
parameter (which is related to the magnetocaloric effect), focusing on the
low-temperature behavior at the infinite-randomness quantum critical point
and in the associated quantum Griffiths phases.

Our results confirm the predictions of a recent scaling theory \cite{Vojta09}
of the Gr\"{u}neisen parameter and its magnetic analog. Specifically, we have found
that the magnetic Gr\"{u}neisen parameter diverges as $\ln(T_0/T)$ with decreasing
temperature in the quantum Griffiths phases on both sides of the transition.
The amplitude of the logarithmic temperature dependence is universally given by
$1/(H-H_c)$ and does not depend on details of the Hamiltonian. The behavior right at
criticality differs from the generic scaling result of Ref.\ \cite{Vojta09}.
This is caused by the self-duality of the random transverse-field Ising model at
its quantum critical point which results in a scaling function that is even in
the distance $r$ from criticality.
Therefore, the leading scaling contribution to the Gr\"{u}neisen parameter
vanishes right at criticality. The next-to-leading term produces a sign change with a
slope that diverges as $[\ln(T_0/T)]^3$ with decreasing temperature. In an experimental
field sweep across the quantum critical point, this rapid sign-change leads a strong feature
in the magnetocaloric effect.

Let us compare the behavior of our random transverse-field Ising chain to that of a clean
chain \cite{GarstRosch05}. The magnetic Gr\"{u}neisen parameter of the clean chain
does not diverge with decreasing temperature but saturates at sufficiently low
$T$ (even though the saturation value can become arbitrarily large close to
criticality). The critical point of the clean chain has the same self-duality as
that of the random chain. Therefore, analogously to our case, the leading scaling contribution
to the Gr\"{u}neisen parameter (which would be proportional to $T^{-1}$)
vanishes in the quantum critical region. This results in a sign change right at the critical field
with a Gr\"{u}neisen ``susceptibility'' $\chi_{\Gamma_H} \sim T^{-2}$.

The random transverse-field Ising chain studied here provides one of the simplest examples
of a field-tuned quantum phase transition in the presence of disorder. It is important to
note that the model realizes random-$T_c$ (random mass) disorder. This means the
disorder does not (locally) break the order parameter symmetry, it just favors
one phase over the other. In many realistic disordered systems, the interplay of an
external (uniform) magnetic field and impurities and defects can generate stronger
random-field type effects \cite{FishmanAharony79,AnfusoRosch09}. In particular, this
applies to LiHo$_x$Y$_{1-x}$F$_4$ which is reasonably well approximated by a transverse-field
Ising model otherwise \cite{TGKSF06,Schechter08}.
If strong enough, such random fields invalidate the infinite-randomness critical point
scenario and require a separate analysis.

\begin{acknowledgement}
This work has been supported in part by the NSF under grants no. DMR-0339147 and
DMR-0506953, by Research Corporation, and by the University of Missouri Research Board.
\end{acknowledgement}

% Use the following code if you wish to generate your bibliography with BibTeX;
% replace the string "pss-demo" below with the name(s) of
% the BibTeX data base(s) you want to use.
% The resulting bibliography-output (the contents of the .bbl file)
% must be pasted back into this file before submission.
%
\bibliographystyle{pss}
\bibliography{../00Bibtex/rareregions}

\end{document}